\documentclass[12pt]{iopart}
\usepackage{graphicx,subfigure}

\usepackage{iopams}

\begin{document}

\title{Study on high pressure plasma produced by ArF laser}

\author{Norio Tsuda$^*$ and Jun Yamada}

\address{Dept. Electronics, Aichi Institute of Technology,\par
 1247 Yakusa-cho, Yachigusa, Toyota, 4700392, JAPAN\par
\textit{* E-mail:n-tsuda@aitech.ac.jp}}

\begin{abstract}

When an ArF excimer laser beam was focused in a high pressure argon gas from 50 to 130 $atm$, the plasma development is observed by streak camera from side window of chamber. The high pressure ArF laser plasma develops symmetrically and the plasma produced by ArF excimer laser hardly develops as compared with the plasma produced by XeCl. The photon energy of ArF laser light is higher than the XeCl laser. The transmittance of ArF laser light was measured. Almost all the laser light is transmitted as the frequency of laser light is higher than the plasma frequency, and the energy of laser light is hardly working for plasma development. The backward and forward plasma development produced by the ArF laser was calculated by the breakdown wave and the radiation supported shock wave, which agreed with the experimental one. The backward development mechanism of high pressure ArF plasma is same as one of plasma produced by XeCl laser. However, the forward development mechanism of ArF plasma differs from one of plasma produced by XeCl laser.

\end{abstract}




\section{Introduction}

\hspace{10mm}When a high pressure gas is irradiated by a focused laser light, a hot and dense plasma is produced.\cite{1} The various studies have been carried out on the mechanism of the gas breakdown,\cite{2} the expansion of the plasma, and the interaction between the laser light and the plasma.\cite{3}

With the development of an excimer laser, a powerful ultraviolet laser light can be easily obtained. As a photon energy of a ultraviolet light is higher than that of a visible light, it is expected that the dense plasma could be efficiently produced by the excimer laser. When a XeCl excimer laser was focused in high pressure argon gases, the dense plasma developed not only backward but also forward,\cite{1} which is different from one produced by visible laser light. However the plasma development dependence on the wavelength  has not been studied yet.

When the ArF laser light is irradiated in high pressure Ar gas up to 130 $atm$ whose wavelength is shorter than the XeCl excimer, the development mechanism of laser plasma has been investigated.

\section{Experimental arrangement}

\hspace{10mm}The experimental arrangement is shown in Fig. 1. The ArF excimer laser with a maximum power of 25 $MW$ , a wavelength of 192 $nm$ and a full half-width of 15 $ns$ is focused.  The output power of laser radiation is controlled by an optical filter, and the laser light is focused at the center of the high pressure gas chamber by a lens with a focal length of 40 $mm$ . The waveform of ArF laser pulse is shown in Fig.2. As the laser light is a rectangle of 11 $mm$ ~24 $mm$ , the focused laser light at the focal spot makes an ellipse of 120 $\mu m$ ~ 80 $\mu m$. The pressure ranges from 50 to 130 $atm$. The dynamic behavior of laser induced plasma is observed by a streak camera. The streak image is displayed by a dummy color of light intensity on a monitor. The plasma boundary is determined by a threshold intensity and the plasma boundary is drawn by a plotter. 

\section{Experimental Results}
\subsection{Streak images}

\hspace{10mm}The streak images of plasma produced by XeCl excimer laser are shown in Fig. 3a. The laser light is irradiated from the right, the time is scanned from top to bottom, the horizontal direction shows the plasma length, and the inside of the boundary shows the plasma. The plasma develops not only backward but also forward. The plasma produced by threshold laser power does not almost develop, it develops symmetrically.

The streak images of plasma produced by ArF excimer laser are shown in Fig. 3b. The directions of irradiated laser light and scan are same as XeCl streak images. The laser light almost transmits in the plasma, because the frequency of laser light may be higher than the plasma frequency. The plasma does not almost develop, it develops symmetrically. 

\subsection{Transmitted waveform}

\hspace{10mm}The backward plasma length of ArF laser plasma is shorter than one of XeCl laser plasma. 
The incident waveform and the transmitted waveform of ArF laser pulse are measured, shown in Fig. 4. The XeCl laser light is almost absorbed by the plasma after the plasma is produced. However, the ArF laser light is hardly decayed at all, and the absorption coefficient is under 20 $\% $ . The frequency of laser light may be higher than the plasma frequency. 

\subsection{Calculated backward plasma development}

\hspace{10mm}The backward developments of ArF laser plasma are calculated using the breakdown wave and radiation supported shock wave. They are same development mechanisum as XeCl laser plasma. However, in this calculation, the decrease of energy absored by plasma is taken into account. The red lines show the calculated backward plasma development in Fig. 5. They agree with the experimental results well. 

\section{Conclusion}

\hspace{10mm}When the high pressure argon gas is irradiated by the focused ultraviolet laser light, a hot and dense plasma is produced at the focal spot. The dynamic behavior of the laser produced-plasma is observed by the streak camera. 
The plasma produced by XeCl excimer laser synmetrically develops not only backward but also forward. On the other hand, the ArF plasma develops synmetrically and the plasma hardly develops, and the laser light is hardly absorbed backward plasma surface. The backward development of ArF laser plasma is calculated by same development mechanisum as one of XeCl laser plasma, which is taking into account the energy absorbed by plasma surface. The calculated backward plasma development agrees with the experimental one.

\begin{flushleft}
REFERENCE
\end{flushleft}

\begin{figure}[htbp]
  \begin{tabular}{cc}
   \begin{minipage}{1.0\textwidth}
    \begin{center}
     \includegraphics[scale=0.5]{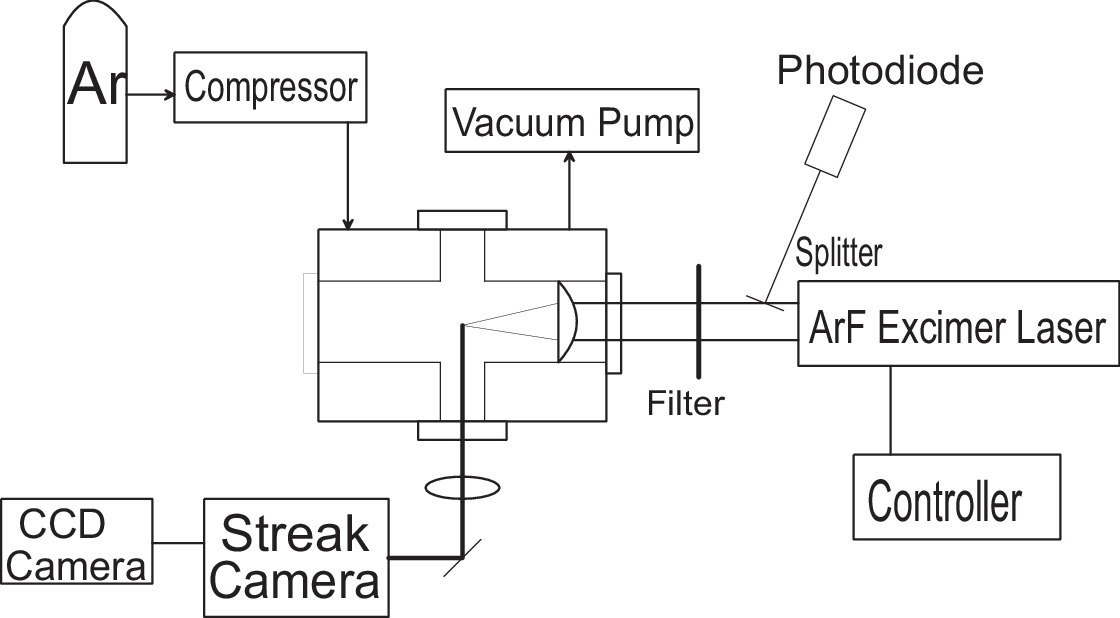}
     \caption{Experimental arrangement}
     \label{label 1}
    \end{center}
   \end{minipage}
\vspace{10mm}
   \\
   \begin{minipage}{1.0\textwidth}
    \begin{center}
     \includegraphics[scale=0.5]{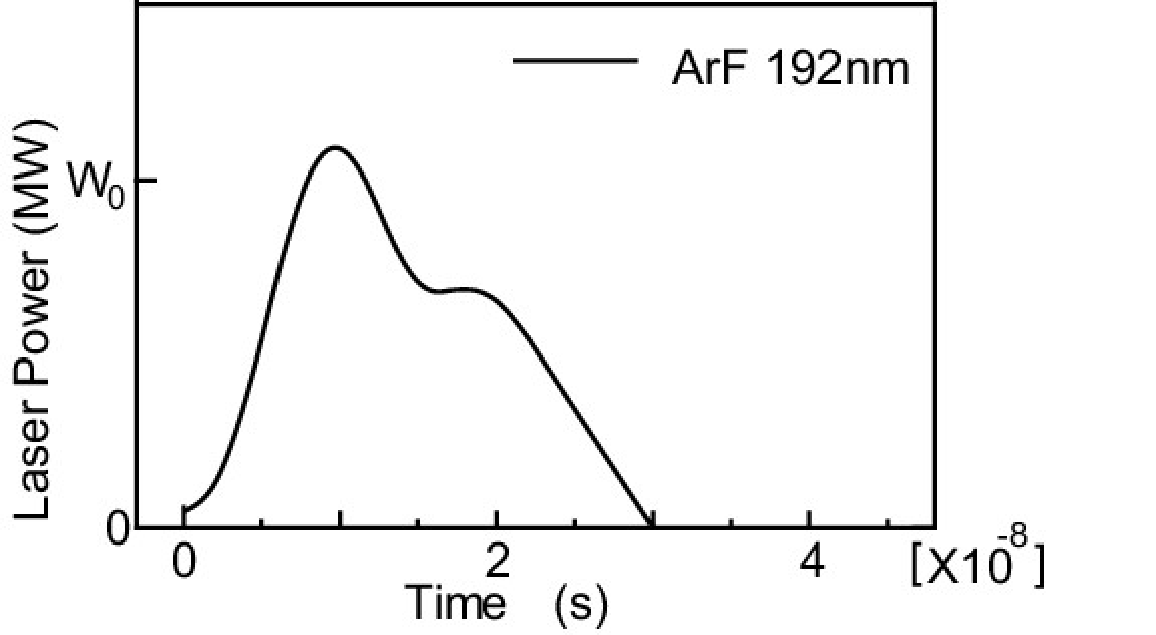}
     \caption{Laser pulse}
     \label{label 2}
    \end{center}
   \end{minipage}
  \end{tabular}
\end{figure}

\begin{figure}[hbp]
 \begin{center}
  \begin{tabular}{c}
  \subfigure[XeCl laser plasma streak images]{\includegraphics[width=0.63\linewidth]{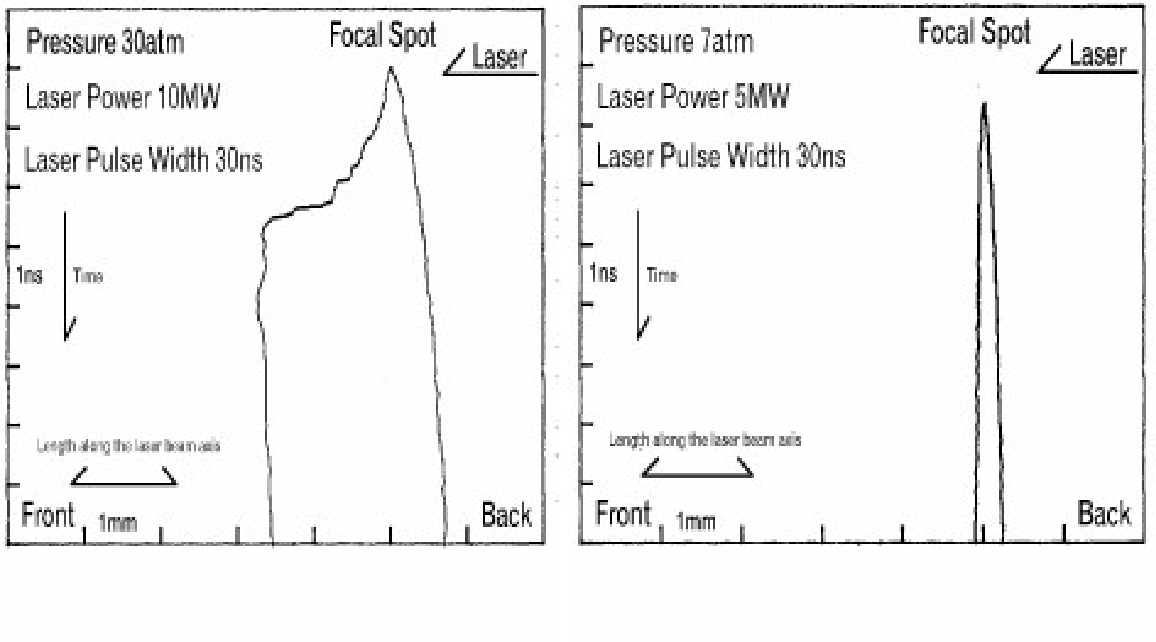}}\\
  \subfigure[ArF laser plasma streak images]{\includegraphics[width=0.66\linewidth]{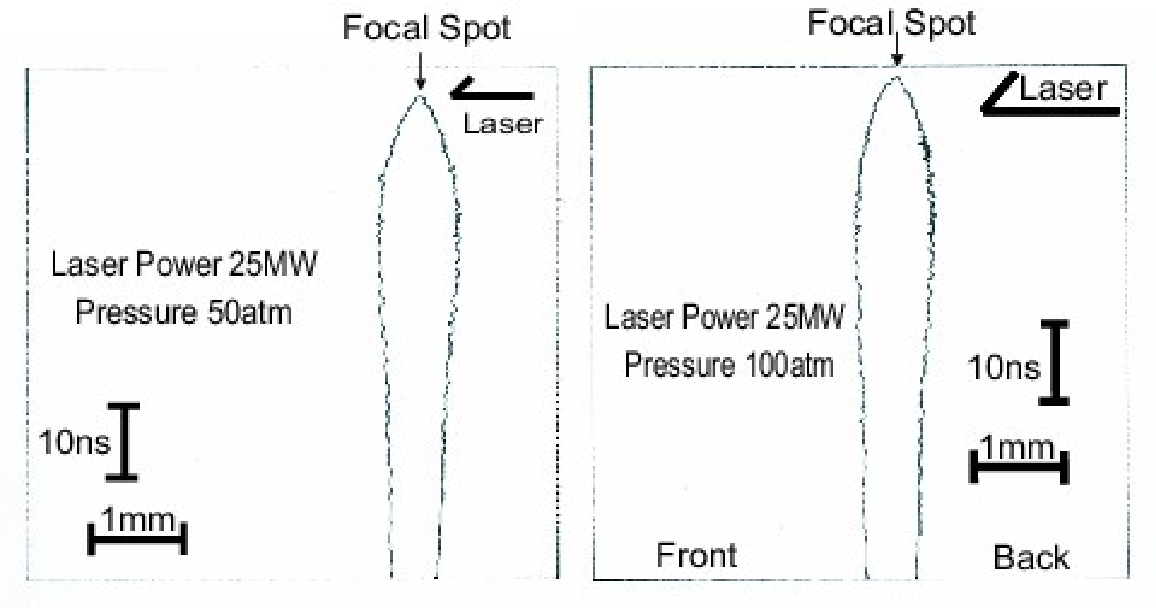}}
  \end{tabular}
    \caption{Streak images}
  \end{center}
 \label{fig:3}
\end{figure}

\begin{figure}[htbp]
  \begin{tabular}{cc}
   \begin{minipage}{0.4\textwidth}
    \begin{center}
     \includegraphics[scale=0.4]{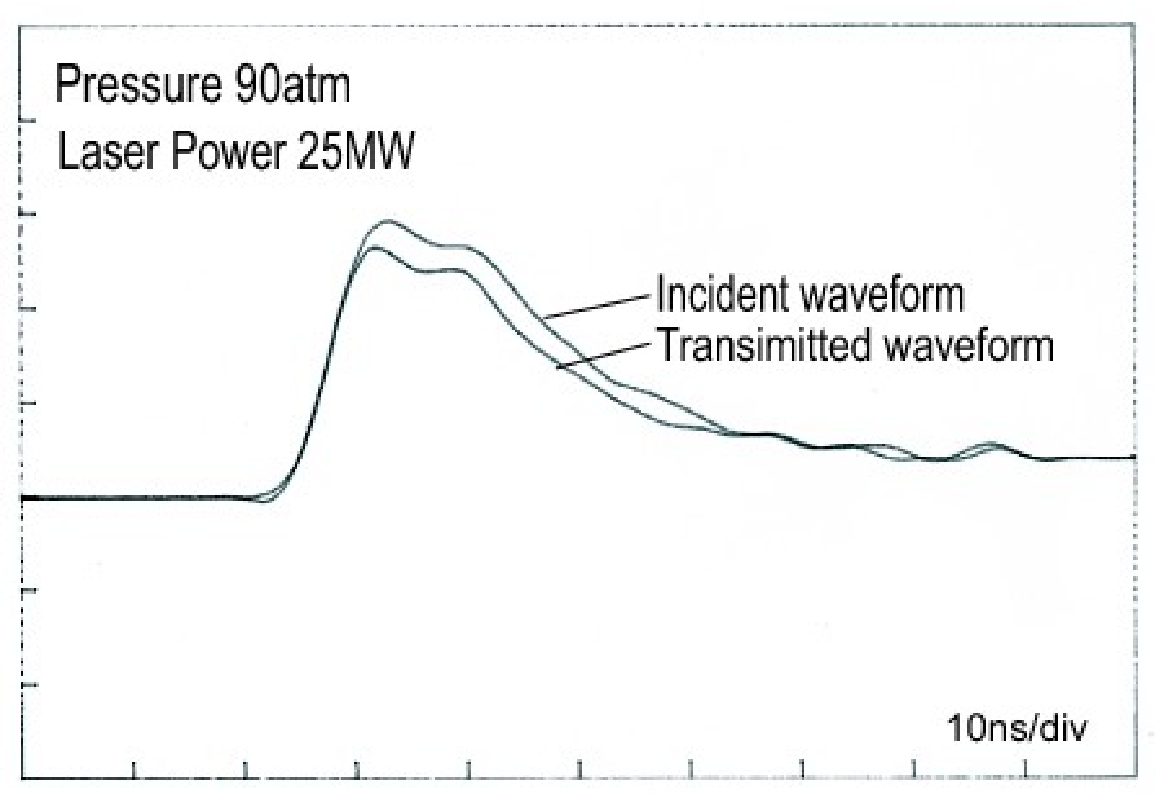}
     \caption{ArF laser pulse and transmitted waveform}
     \label{label 4}
    \end{center}
   \end{minipage}
   \begin{minipage}{0.7\textwidth}
    \begin{center}
     \includegraphics[scale=0.32]{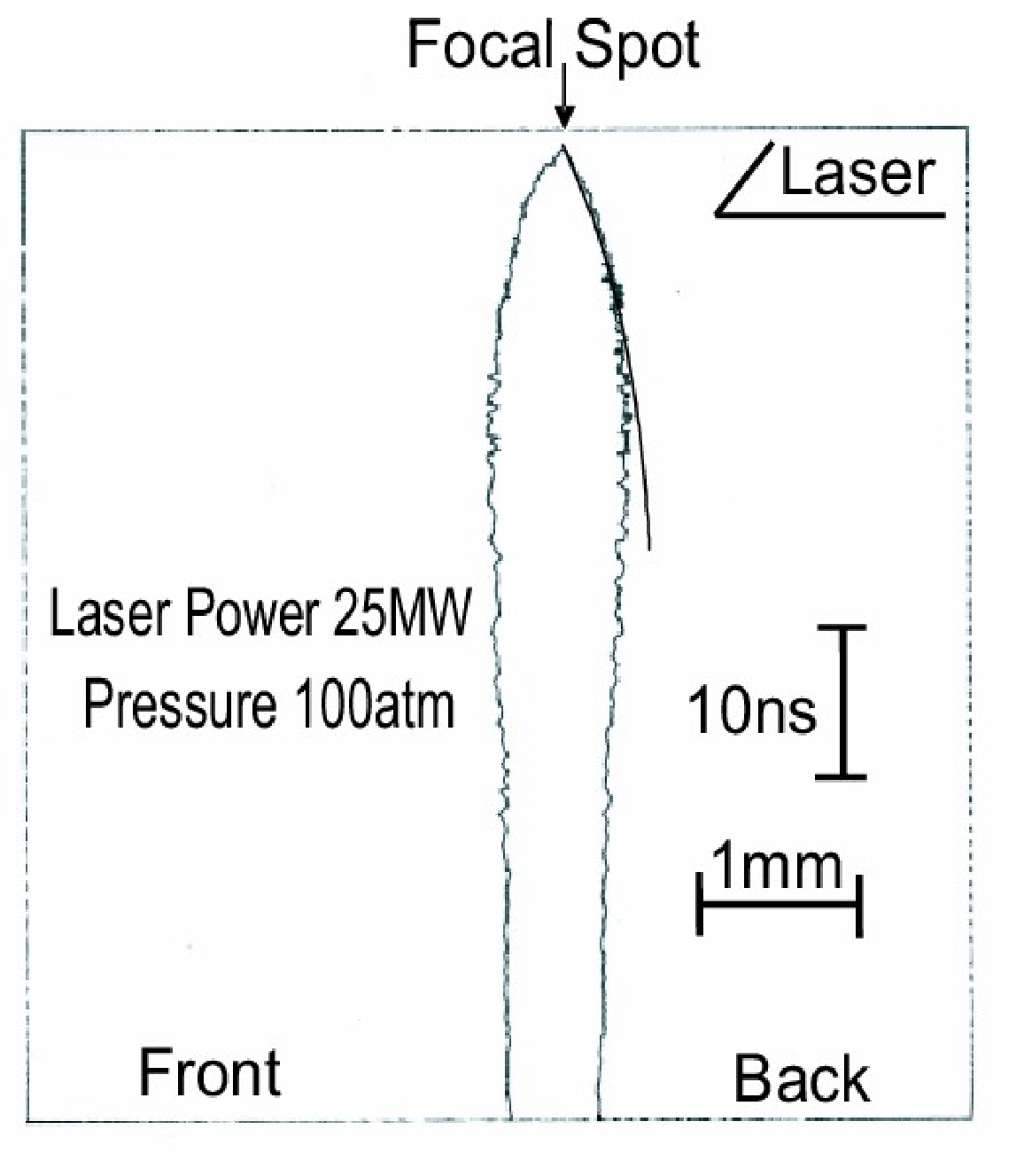}
     \caption{Calculated backward plasma development}
     \label{label 5}
    \end{center}
   \end{minipage}
  \end{tabular}
\end{figure}


\begin{thebibliography}{4}
\bibitem{1}N. Tsuda et al. J. Appl. Phys. 81 582 (1997).
\bibitem{2}R.G. Meyerand et al. Phys. Rev. Lett. 11 401 (1963).
\bibitem{3}G. V. Ostrovskaya et al. Sov. Phys. Usp. 16 834 (1974).
\end{thebibliography}
\end{document}